\documentstyle[graphicx,aps,multicol]{revtex}
\draft

\begin{document}

\title{Determination of the electromagnetic character of soft dipole modes
solely based on quasicontinuous $\gamma$ spectroscopy}
\author{A.~Voinov\footnote{Electronic address: voinov@nf.jinr.ru}}
\address{Frank Laboratory of Neutron Physics, Joint Institute of Nuclear
Research, 141980 Dubna, Moscow region, Russia}
\author{A.~Schiller\footnote{Electronic address: Andreas.Schiller@llnl.gov}}
\address{Lawrence Livermore National Laboratory, L-414, 7000 East Avenue,
Livermore CA-94551}
\author{M.~Guttormsen, J.~Rekstad, and S.~Siem}
\address{Department of Physics, University of Oslo, N-0316 Oslo, Norway}
\maketitle

\begin{abstract}
We show that the combined analysis of the quasicontinuous $\gamma$ spectra from
the $(^3$He,$\alpha)$ and the $(n_{\mathrm{th}},2\gamma)$ reactions gives the
possibility to measure the electromagnetic character of soft dipole resonances.
Two-step $\gamma$-cascade spectra have been calculated, using level densities
and radiative strength functions from the $(^3$He,$\alpha\gamma)$ reaction. The
calculations show that the intensity of the two-step cascades depends on the
electromagnetic character of the soft dipole resonance under study. The
difference reaches 40-100\% which can be measured experimentally.
\end{abstract}


\begin{multicols}{2}

\section{Introduction}

A soft resonance was observed for the first time by Jackson and Kinsey
\cite{JK51} in the $(n,\gamma)$ reaction for heavy nuclei $(A>70)$ near closed
shells as an 'anomalous' bump appearing in the quasicontinuous $\gamma$
spectrum at $E_\gamma\sim 5.5$~MeV\@. This resonance was called pygmy resonance
(PR) because of its small strength compared to the known, giant dipole
resonance observed in photonuclear reactions \cite{DB88}. Later, a similar
'anomalous' bump was observed for the well-deformed rare-earth nucleus
$^{170}$Tm, but at the considerably lower energy of $E_\gamma\sim 3.0$~MeV
\cite{JD79}. A systematic investigation of the PR in the rare-earth mass region
and beyond was made by Igashira \sl et al.\rm \cite{IK86}, where they found
that the centroid of the PR is approximately proportional to the mass number
$A$ and varying from as low as 1.5~MeV for $^{142}$Pr to 5.6~MeV for
$^{198}$Au. While all those studies have been performed using radiative neutron
capture, the PR was recently observed also in the radiative strength function
(RSF) derived from the $(^3$He,$\alpha\gamma)$ reaction \cite{VG01}. This
supports the assumption that the fundamental origin of this phenomenon is
independent of the type of nuclear reaction.

However, in spite of the fact that the PR has been studied for a long time, the
type of nuclear excitation responsible for the PR in the RSF remains unknown.
Theory \cite{La71} predicts an $E1$ PR around $5.5$~MeV in $^{208}$Pb. The
origin for the PR in this calculation is a supposed concentration of electric
dipole strength in few low-energy states composed of $3p^{-1}4s$, $3p^{-1}3d$,
and $2f^{-1}3d$ neutron configurations which are relatively weakly coupled to
the giant electric dipole resonance (GEDR). This model suggest a quite strong
PR which is not supported by experimental data. However, recently some
indications in $(\gamma,\gamma')$ experiments for the existence of a weak $E1$
PR in lead nuclei have been discussed \cite{EB00}. On the other hand, the
hypothesis has been suggested \cite{SG00a} that the PR observed at
$E_\gamma\sim 3.0$~MeV in the quasicontinuous $\gamma$ spectra of well-deformed
rare-earth nuclei and the so-called `scissors mode' excitation studied by the
nuclear resonance fluorescence method (NRF) \cite{PB98} have the same physical
origin. This would imply $M1$ multipolarity for the PR and is supported by the
similarity of the measured resonance parameters in the quasicontinuous $\gamma$
spectrum on the one hand, and the averaged properties of individual $M1$
transitions seen in NRF experiments on the other hand. Finally, one can not
exclude the possibility that the multipolarity of the PR can be different for
different nuclei, i.e., one has to deal with different phenomena in different
mass or energy regions. In conclusion, while the commonly-called pygmy
resonance is most probably a \em dipole \rm resonance, the \em electromagnetic
character \rm of this soft dipole mode remains unknown.

In NRF experiments, one can, in principle, measure the multipolarity and
character of the involved $\gamma$ transitions, but one is usually limited to
the investigation of the soft dipole resonances built on the ground state. Even
so, for reliable results above $\sim 4$~MeV, one should prefer the use of a
polarized photon beam over common Compton polarimeters \cite{PB02}. Anyway, in
order to study average properties of the soft dipole modes built on excited
states, one has to analyze quasicontinuous $\gamma$ spectra. To do so using a
polarimeter might prove very hard, since the soft dipole mode is in general a
weak resonance superimposed on the rather large tail of the GEDR \cite{VG01}
giving a very low signal-to-background ratio in the analysis. Unfortunately,
with the exclusion of Compton polarimetry, there are no other known direct
experimental methods\footnote{The involved transition energies are in general
too high for an efficient use of conversion electrons for multipolarity and
character determination.} to determine the electromagnetic character of soft
dipole modes in quasicontinuous $\gamma$ spectroscopy. In this work, we propose
the possibility to determine the character of soft dipole modes solely by using
quasicontinuous $\gamma$ spectra from two different types of reactions, namely
the $(^3$He,$\alpha\gamma)$ and the $(n,2\gamma)$ reactions applied to the
study of the same nucleus. This will enable us to study soft dipole modes built
on excited states and their response to finite nuclear temperature.

\section{$(^3$He,$\alpha\gamma)$ reaction}

The $(^3$He,$\alpha\gamma)$ reaction is used for the extraction of level 
densities and RSFs from primary $\gamma$ spectra $P(E_i,E_{\gamma})$ for
excitation energies $E_i$ between 0 and the neutron binding energy $B_n$ and
for spins between 2 and 6$\hbar$. The fundamental assumption behind the
extraction procedure is the Brink-Axel hypothesis \cite{Br55,Ax62} which
implies that the probability of radiative decay in the statistical regime,
represented by the primary $\gamma$ spectra, can be expressed simply as a
product of the final-state level density $\rho(E_f=E_i-E_\gamma)$ and the
$\gamma$-ray transmission coefficient $T(E_{\gamma})$
\begin{equation}
P(E_i,E_{\gamma})\propto T(E_{\gamma})\rho(E_f).
\end{equation}
The $\gamma$-ray transmission coefficient is now proportional to
$\sum_{XL}E_{\gamma}^{2L+1}f^{XL}(E_{\gamma})$, where $f^{XL}(E_{\gamma})$ is
the RSF of the multipolarity $XL$. The details of the experimental extraction
procedure have been described in Refs.\ \cite{HB95,SB00} and references
therein.

In Ref.\ \cite{VG01}, the total RSFs has been obtained from such experiments
and analyzed for the $^{161,162}$Dy and $^{171,172}$Yb nuclei. It has been
established that the total RSF can be readily described by the sum of
\begin{itemize}
\item the tail of the GEDR as given by the Kadmenski{\u{\i}}-Markushev-Furman
(KMF) model \cite{KM83} with a fixed nuclear temperature fitted to the
experimental data
\item the Lorentzian-shaped, single-humped spin-flip giant magnetic dipole
resonance (GMDR) located around 7~MeV of excitation energy
\item a Lorentzian-shaped soft dipole mode or PR, where the resonance
parameters were determined from experiment \cite{VG01}.
\end{itemize}
Although the centroid, width and strength of the soft dipole mode could be
determined from experiment, its electromagnetic character could not. Thus,
the question still remains open, whether it is of $E1$ or $M1$ type. The
remaining of this work is therefore devoted to describing a method of 
determining the character of soft dipole modes observed in quasicontinuous
$\gamma$ spectra.

\section{$(n,2\gamma)$ reaction}

In the $(n,2\gamma)$ reaction with thermal neutrons, the compound state $i$ is
formed with spin and parity $J_i^\pi=I^\pi\pm 1/2$, where $I^\pi$ is the spin
and parity of the target nucleus. The most intense two-step $\gamma$ cascades
(TSCs) accessible to experimental study populate low-lying excited states with
excitation energies $E_{f}$ up to around 1~MeV \cite{VS00}.

Thus, TSCs start from the initial (compound) state $i$ with a certain spin and
parity to a final state $f$ through all possible intermediate levels $j$ with
excitation energies from $E_{f}$ to $E_i=B_n$. The possible spins of the
intermediate levels $J_j^{\pi}$ are determined by the multipolarities in the
TSCs. Assuming that the most intense TSCs have dipole multipolarity, the most
probable spin interval for the intermediate level can be defined as
\begin{equation}
\max\left[J_i-1,J_f-1\right]\leq J_j\leq\min\left[J_i+1,J_f+1\right].
\end{equation}

Assuming the validity of the statistical theory, the intensity of TSCs between
$i\rightarrow j\rightarrow f$ levels is determined by the partial and total
radiative widths of both, the initial $\Gamma_{ij}$,$\Gamma_i$ and intermediate
$\Gamma_{jf}$,$\Gamma_{j}$ levels, respectively
\begin{equation}
I^{\gamma\gamma}_{ijf}=\frac{\Gamma_{ij}}{\Gamma_i}\
\frac{\Gamma_{jf}}{\Gamma_j}.
\end{equation}
Partial radiative widths of any arbitrary level $l$ can be expressed in terms
of the RSF $f^{XL}(E_\gamma)$
\begin{equation}
\Gamma^{XL}_{lm}(E_\gamma=E_l-E_m)=f^{XL}(E_\gamma)E_\gamma^{(2L+1)}D_l(E_l),
\end{equation}
where $D_l$ is the energy spacing of levels having the same spin and parity as
$l$ \cite{BE73}. The total radiative width is determined by the sum of all
partial radiative widths $\Gamma_{lm}$ summed over all $\gamma$ transitions
populating levels $m$ lying below $l$ in excitation energy, i.e.,
$\Gamma_l=\sum_{m<l}{\Gamma_{lm}}$.

The multipolarities of the transitions in a TSC depend on the spins and
parities of the connected states. Here, we assume that the main
contributions come from $E1$ and $M1$ $\gamma$ transitions. Also some $E2$
strength will be present, but is neglected in the present discussion. Thus,
if the parities of the initial $i$ and final $f$ states are different, the
$(E1,M1)$ and $(M1,E1)$ sequences of transition multipolarities are
possible. In the case of equal parities of the initial $i$ and final $f$
states, the $\gamma$-transitions of the TSCs have $(E1,E1)$ and $(M1,M1)$
multipolarities. Because there is in general no possibility to distinguish
between primary $E_{\gamma_1}$ and secondary $E_{\gamma_2}$ transitions in
the cascade,\footnote{Ordering the $\gamma$ transitions in TSC experiments
has been attempted in, e.g., \cite{AK94}} the observed spectrum with the
total TSC energy $E_{\mathrm{tot}}=E_{\gamma_1}+E_{\gamma_2}=B_n-E_f$ is
symmetric with respect to the energy $E_{\gamma}=E_{\mathrm{tot}}/2$ and is
determined by
\begin{eqnarray}
\lefteqn{\frac{{\mathrm d}N(E_\gamma;E_{\mathrm{tot}})}{{\mathrm d}E_\gamma}=}
\label{eq:spectrum}\\
&&\sum_{\stackrel{\scriptstyle L,L',J}{\scriptstyle X,X',\pi}}
\frac{\Gamma^{XL}_{ij}(E_{\gamma})}{\Gamma_i}\
\frac{\Gamma^{X'L'}_{jf}(E_{\mathrm{tot}}-E_{\gamma})}{\Gamma_j}
\rho^{J^\pi}_j(B_n-E_\gamma)
\nonumber\\
&&+\sum_{\stackrel{\scriptstyle L,L',J}{\scriptstyle X,X',\pi}}
\frac{\Gamma^{XL}_{ih}(E_{\mathrm{tot}}-E_{\gamma})}{\Gamma_i}\
\frac{\Gamma^{X'L'}_{hf}(E_{\gamma})}{\Gamma_h}
\rho^{J^\pi}_h(E_f+E_\gamma),\nonumber
\end{eqnarray}
where $\rho_j$ and $\rho_h$ denote the density of intermediate levels populated
by TSCs, and the sums run over all possible combinations of $L,L',J$ and
$X,X',\pi$. Thus, we notice that the quasicontinuous $\gamma$-spectra from
both, the $(^3$He,$\alpha\gamma)$ and the $(n,2\gamma)$ reaction are determined
by the same level densities and RSFs. This allows us to use the quantities
which can be extracted from the $(^3$He,$\alpha\gamma)$ reaction in order to
calculate the quasicontinuous spectra of TSCs following thermal neutron
capture. In the following, we will mainly focus on the influence of the
unknown electromagnetic character of the soft dipole mode on calculated TSC
spectra.

\section{Character of soft dipole modes}

For the sake of simplicity, we first consider the TSC with the $(E1,M1)$
sequence of transition multipolarities. If the soft dipole mode is of $M1$
character, the corresponding strength function $f^{\mathrm{SD}}$ should be
added incoherently to the spin-flip GMDR strength function $f^{\mathrm{SF}}$
giving the total magnetic dipole strength function
$f^{M1}=f^{\mathrm{SF}}+f^{\mathrm{SD}}$. The same is correct for the radiative
widths $\Gamma^{M1}=\Gamma^{\mathrm{SF}}+\Gamma^{\mathrm{SD}}$. The $E1$
strength function is then completely determined by the tail of the GEDR
(modeled in this work by the KMF approach). The intensity of $(E1,M1)$ TSCs can
therefore be written as
\begin{equation}
I^{{\mathrm{SD}}=M1}_{ijf}=\frac{\Gamma^{\mathrm{KMF}}_{ij}(E_{\gamma_1})}
{\Gamma_i}\
\frac{[\Gamma^{\mathrm{SF}}_{jf}(E_{\gamma_2})+\Gamma^{\mathrm{SD}}_{jf}
(E_{\gamma_2})]}{\Gamma_j}
\label{eq:intcas1}
\end{equation}
or
\begin{equation}
I^{{\mathrm{SD}}=M1}_{ijf}=\frac{\Gamma^{\mathrm{KMF}}_{ij}(E_{\gamma_1})}
{\Gamma_i}\
\frac{\Gamma^{\mathrm{SF}}_{jf}(E_{\gamma_2})}{\Gamma_j}
\left[1+\frac{\Gamma^{\mathrm{SD}}_{jf}(E_{\gamma_2})}
{\Gamma^{\mathrm{SF}}_{jf}(E_{\gamma_2})}\right].
\label{eq:intcas2}
\end{equation}

In the case of the soft dipole mode being of $E1$ character, the total electric
dipole strength function $f^{E1}$ is given by the incoherent sum of the KMF
strength function $f^{\mathrm{KMF}}$ and the soft dipole resonance
$f^{\mathrm{SD}}$, i.e., $f^{E1}=f^{\mathrm{KMF}}+f^{\mathrm{SD}}$. The
intensity of the $(E1,M1)$ cascade can then be expressed as
\begin{equation}
I^{{\mathrm{SD}}=E1}_{ijf}=\frac{\Gamma^{\mathrm{KMF}}_{ij}(E_{\gamma_1})}
{\Gamma_i}\
\frac{\Gamma^{\mathrm{SF}}_{jf}(E_{\gamma_2})}{\Gamma_j}
\left[1+\frac{\Gamma^{\mathrm{SD}}_{ij}(E_{\gamma_1})}
{\Gamma^{\mathrm{KMF}}_{ij}(E_{\gamma_1})}\right].
\label{eq:intcas3}
\end{equation}
The ratio of the cascade intensities (\ref{eq:intcas2}) and (\ref{eq:intcas3})
is finally determined by
\begin{equation}
\frac{I^{{\mathrm{SD}}=E1}_{ijf}}{I^{{\mathrm{SD}}=M1}_{ijf}}=
\frac{\left[1+\frac{\Gamma^{\mathrm{SD}}_{ij}(E_{\gamma_1})}
{\Gamma^{\mathrm{KMF}}_{ij}(E_{\gamma_1})}\right]}
{\left[1+\frac{\Gamma^{\mathrm{SD}}_{jf}(E_{\gamma_2})}
{\Gamma^{\mathrm{SF}}_{jf}(E_{\gamma_2})}\right]}.
\label{eq:intcas4}
\end{equation}

One can see that this ratio depends on the ratios of the soft dipole and the
KMF-GEDR strength function and on the ratio of the soft dipole and the
spin-flip GMDR strength function for primary ($i\rightarrow j,\ E_{\gamma_1}$)
and secondary ($j\rightarrow f,\ E_{\gamma_2}$) transitions, respectively. It
is known from Ref.\ \cite{VG01} that the spin-flip GMDR strength function
$f^{\mathrm{SF}}$ is less than $\sim 20$\% in magnitude than the tail of the
GEDR strength function $f^{\mathrm{KMF}}$ according to the KMF model.
Therefore, it is expected that the intensity of the calculated TSCs will be
higher in the case of $M1$ character of the soft dipole resonance under study
than for $E1$ character, provided that the parities of the initial and final
state are opposite. The main differences in the calculated TSC spectra for the
two possible characters of the soft dipole mode are expected to be located
around its centroid, i.e., in the region of $E_\gamma\sim 2.5-3.5$~MeV for
deformed rare earth nuclei.

By analogy with Eqs.\ (\ref{eq:intcas2}) and (\ref{eq:intcas3}), similar
expressions can be written for the $(M1,E1)$ sequence of multipolarities in the
TSC\@. For the $(E1,E1)$ and the $(M1,M1)$ sequences, i.e., for equal parities
of the initial and final state, the situation is more complex and it is more
difficult to draw conclusions from simple relations similar to Eqs.\
(\ref{eq:intcas1}-\ref{eq:intcas4}). In general, one can state that in these
cases, the calculated TSC intensities will be higher for $E1$ than for $M1$
multipolarity of the soft dipole mode under study, but the differences between
the two calculations are smaller than in the case of opposite parities of the
initial and final state.

In order to investigate the intensity difference of calculated TSC spectra due
to the two possible characters of the soft dipole resonance, we have calculated
such spectra according to Eq.\ (\ref{eq:spectrum}) for the two even-even nuclei
$^{162}$Dy and $^{172}$Yb and the odd nucleus $^{161}$Dy. For all of these
nuclei, the level density and total RSF, as well as the resonance parameters of
the soft dipole mode have been obtained from $(^3$He,$\alpha\gamma)$
experiments \cite{VG01}. We will investigate TSCs populating the first $2^+$
level of the $^{162}$Dy and $^{172}$Yb nuclei. The initial states (which are
formed at $B_n$) have mainly spin and parity $J_i^\pi=3^+$ and $J_i^\pi=0^-$
for the $^{162}$Dy and $^{172}$Yb nuclei, respectively. Thus, we can
investigate the $(E1,E1)$ and $(M1,M1)$ sequences of TSCs in the case of the
$^{162}$Dy nucleus and the $(E1,M1)$ and $(M1,E1)$ sequences in the case of the
$^{172}$Yb nucleus. The $^{161}$Dy nucleus is interesting because it has two
low lying levels $5/2^+$ and $5/2^-$ which are very close in excitation energy
(0 and 25.6~keV respectively) and which have the same spin but opposite parity.
The initial state (formed by thermal neutron capture) has $J_i^\pi=1/2^+$.
Thus, we can investigate $(E1,E1)$ and $(M1,M1)$ as well as $(E1,M1)$ and
$(M1,E1)$ cascades for the same nucleus.

Because the spin interval accessible for $\gamma$-transitions is different for
the $(^3$He,$\alpha\gamma)$ and $(n,2\gamma)$ reaction, a spin distribution of
the level density has to be taken into account. We assume the spin distribution
according to \cite{GC65}
\begin{equation}
g(E,I)=\frac{2I+1}{2\sigma^2}\exp\left[-(I+1/2)^2/2\sigma^2\right],
\end{equation}
where $\sigma$ is the excitation-energy dependent spin cut-off parameter.
Furthermore, we assume equal amounts of levels with positive and negative
parities for all relevant spins and excitation energies.

\begin{figure}\centering
\includegraphics[totalheight=17.9cm]{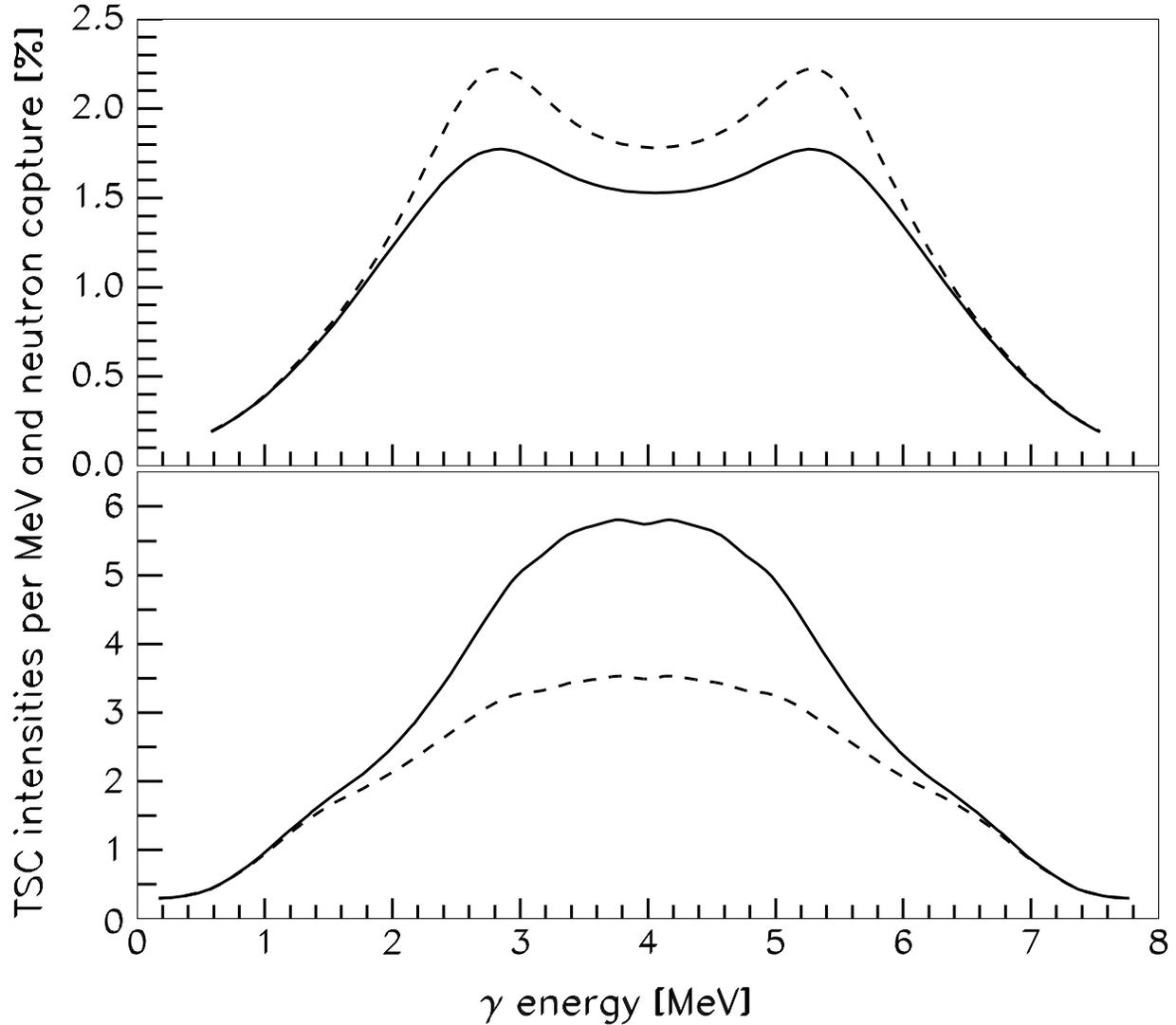}
\caption{Calculated spectra of TSCs to the first $2^+$ state from the 
$^{161}$Dy$(n,2\gamma)^{162}$Dy (upper panel) and 
$^{171}$Yb$(n,2\gamma)^{172}$Yb (lower panel) reactions. Due to the unknown 
multipolarity of the soft dipole mode, we obtain two different solutions 
corresponding to $E1$ (dashed line) or $M1$ (solid line) multipolarity. 
Comparison to experimental data will reveal which of the $E1$ or $M1$ 
assignment is the most probable.}
\label{fig:even}
\end{figure}

\begin{figure}\centering
\includegraphics[totalheight=17.9cm]{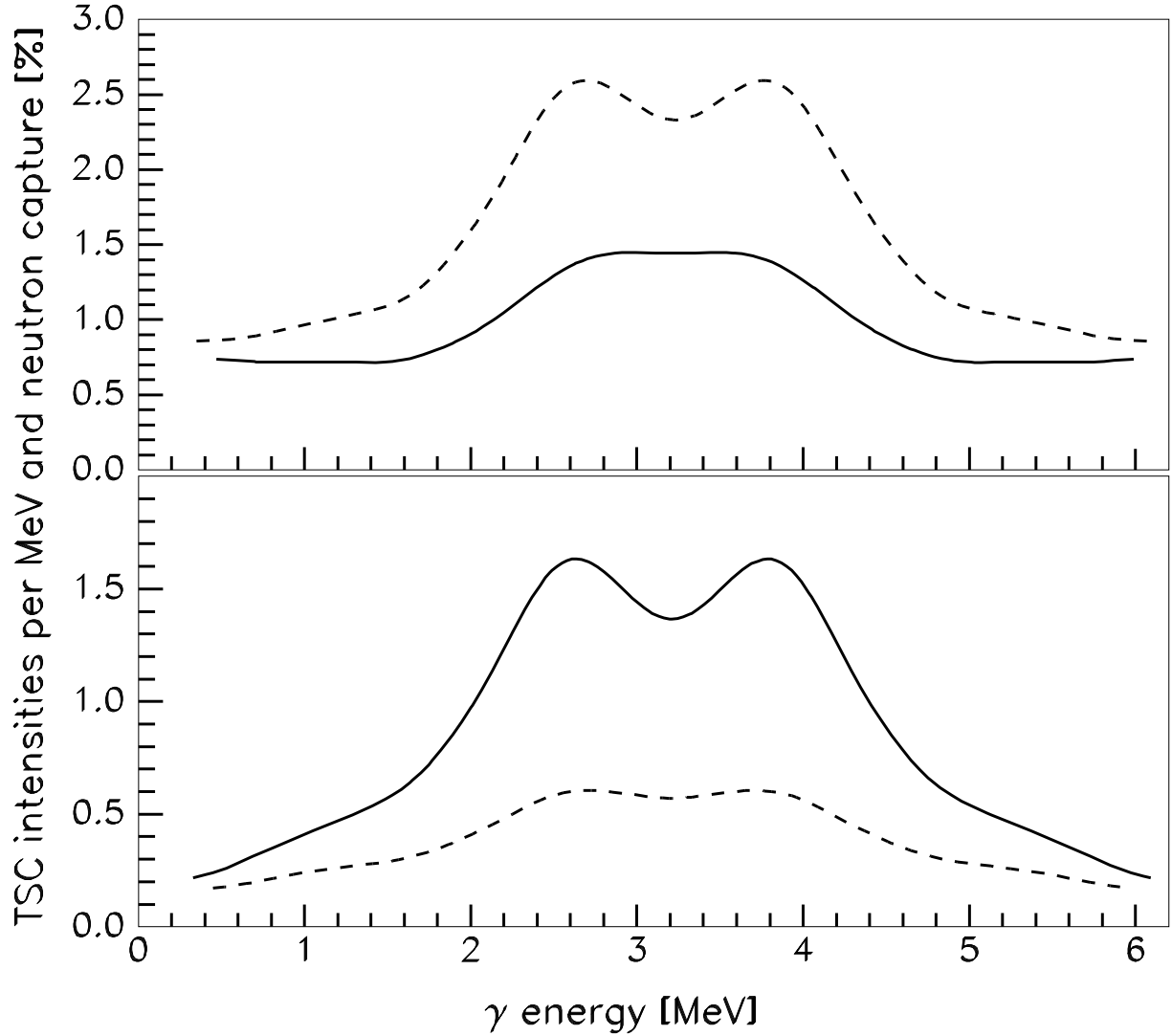}
\caption{Same as Fig.\ \protect{\ref{fig:even}} but for the
$^{160}$Dy$(n,2\gamma)^{161}$Dy reaction and for two different final states
with spins $5/2^+$ (upper panel) and $5/2^-$ (lower panel). The calculations 
must describe experimental TSC spectra to both final spins simultaneously 
assuming one and the same multipolarity for the soft dipole mode.}
\label{fig:odd}
\end{figure}

The calculated TSC spectra for the different nuclei and final states are given
in Figs.\ \ref{fig:even} and \ref{fig:odd}. The calculations were carried out
for both possible characters of the soft dipole mode, giving two solutions. The
summed intensities of all TSCs (the integrals over the TSC spectra) populating
a fixed low lying level are shown in Table \ref{tab:sum}. The results of the
calculations show that the two possible electromagnetic characters of the soft
dipole resonance yield different spectral distributions of TSCs as well as
different summed intensities. In the more favorable cases where the parities of
the initial and final states are opposite, the summed intensity differs between
30\% for $^{172}$Yb and up to a factor of two for $^{161}$Dy with the $M1$
multipolarity for the soft dipole mode resulting in the higher intensities. The
spectral distributions give differences of up to a factor of two to three at
energies which correspond to the centroid of the soft dipole resonance and show
markedly different shapes. For the less favorable cases where the parities of
the initial and final states are equal, the differences in summed intensities
amounts only to 15\% for the $^{162}$Dy nucleus and $\sim$45\% for the
$^{161}$Dy nucleus with the $E1$ multipolarity giving the higher intensities.
The spectral distributions for the two possible electromagnetic characters
differ between 25\% and a factor of two around the centroid of the soft dipole
resonance. Again the shapes of the TSCs spectra show differences as well.

Thus, we see that the intensity of the TSC transitions can depend strongly on
the character of the soft dipole resonance under study. The effect should
certainly be large enough to be measured experimentally, especially for TSCs
with opposite parities of the initial and final states connected by the TSC
transitions. This fact should open the possibility to use the experimental
measurements of TSC spectra to determinate the character of soft dipole modes
observed in the analysis of $(^3$He,$\alpha\gamma)$ reaction data, i.e., using
solely quasicontinuous $\gamma$ spectroscopy.

\section{Possible systematic errors and other considerations}

It should be noted that the statements and conclusions discussed above are
based on some fundamental assumptions. The most important of these is that the
$\gamma$ emission in both reactions discussed is of statistical nature. For the
$(^3$He,$\alpha\gamma)$ reaction, this means that the decay properties of the
ensemble of reaction selected states within a certain energy bin are
independent of whether these states are directly populated by $\alpha$
particles through the nuclear reaction or by $\gamma$-transitions from higher
excited states. This is believed to be approximately fulfilled because the
relatively long life time of the excited states ($\sim 10^{-15}$~s) gives the
nucleus time to thermalize prior to radiative decay. The same is assumed for
the $(n,2\gamma)$ reaction, i.e., it is assumed that the reaction proceeds
through a compound state which has a long life time and that the radiative
decay properties of this compound state are independent of its formation. It is
also assumed for both reactions that the mean values of the radiative widths
$\langle\Gamma_{if}\rangle$ are independent on the structure of the connected
states and are defined by the relevant level spacings and RSFs.

These assumptions have been partly tested for the $^{162}$Dy nucleus
\cite{VG01}. It has been shown that the measured total $\gamma$-ray spectrum
of the $^{161}$Dy$(n,\gamma)^{162}$Dy reaction with keV neutrons can be very
well reproduced by using the level density and RSFs obtained from the
$^{163}$Dy$(^3$He,$\alpha\gamma)^{162}$Dy reaction. Additionally, for the
$^{162}$Dy nucleus, the essentially same level density and RSF have been
obtained from the $^{162}$Dy$(^3$He,$^3$He'$\gamma)^{162}$Dy reaction
\cite{SG00b}. These results strongly support the statistical model for the
radiative decay in these reactions and the correctness of the extraction
procedure for level densities and RSFs in the case of the $^3$He induced
reactions.

A second difficulty in the above outlined method to measure the multipolarity
of soft dipole resonances comes from the fact that the extraction procedure for
level densities and RSFs applied to the $(^3$He,$\alpha\gamma)$ reaction allows
us to get only the total RSF, i.e., the incoherent sum of all multipolarities.
Decomposition of the total RSF into strength functions connected to definite
multipolarities is determined by corresponding models used to describe the $E1$
and $M1$ strength in nuclei. This model dependent analysis can, in principle,
give additional uncertainties to the analysis of TSC spectra. However, it has
been shown \cite{VG01} that if one tests only widely accepted models of $E1$
and $M1$ strength functions in well-deformed rare earth nuclei, the total RSF
extracted from the $(^3$He,$\alpha\gamma)$ reaction allows one to select only
the KMF model with a constant nuclear temperature for the tail of the GEDR, a
single-humped Lorentzian for the spin-flip GMDR
resonance,\footnote{Double-humped Lorentzians as proposed in \cite{Ri95} were
not considered in the decomposition.} and a Lorentzian for the soft dipole
mode. From this point of view, the measurements of TSC spectra can also be
considered as an additional test of this decomposition.

Finally, investigating the same residual nucleus by the $(^3$He,$\alpha\gamma)$
and $(n,2\gamma)$ reaction, requires two target nuclei with neutron number
differing by two units. If we restrict the search to stable targets of
well-deformed rare-earth nuclei, only 22 of such pairs of target nuclei exist
between Gd and Hf.\footnote{The method outlined in this work should be
applicable to any nucleus. However, most of the published
$(^3$He,$\alpha\gamma)$ and $(n,2\gamma)$ experiments have been devoted to
well-deformed rare-earth nuclei.} Unfortunately, until now, there is, to our
knowledge, no published data set from both reactions populating the same
residual nucleus. Thus, the analysis outlined in this work can not be performed
without additional experimental efforts.

\section{Conclusion}

We have shown that a joint experimental investigation of the same product
nucleus by the two different reactions $(^3$He,$\alpha\gamma)$ and
$(n,2\gamma)$, can be used to determine the character of soft dipole resonances
observed in the RSF. The first reaction is used for extracting the level
density and the total RSF of the residual nucleus. The subsequent decomposition
of this total RSF into the tail of the GEDR, the GMDR, and the soft dipole
resonance is model dependent and subject to possible systematic errors. The
second reaction can then be used to determine the character of the soft dipole
resonance via the measurement of TSC intensities. The most suitable TSCs for
this analysis are those with opposite parity of the initial and final level.
The results from the $(n,2\gamma)$ experiment can also be used as an additional
test for the correct decomposition of the total RSF obtained from the
$(^3$He,$\alpha\gamma)$ experiment into modeled electric and magnetic dipole
strength functions. Finally, the exclusive use of quasicontinuous $\gamma$
spectra for determining the character of soft dipole modes gives us, in 
contrast to typical NRF experiments, for the first time the opportunity to 
study soft dipole resonances built on excited states and their response to
finite nuclear temperatures.

\acknowledgments

Part of this work was performed under the auspices of the U.S. Department of
Energy by the University of California, Lawrence Livermore National Laboratory 
under Contract No.\ W-7405-ENG-48. Financial support from the Norwegian 
Research Council (NFR) is gratefully acknowledged. A.V. would like to thank the
staff of the University of Oslo for their warm hospitality while this 
manuscript was prepared and acknowledges support from a NATO Science Fellowship
under project number 150027/432 given through the Norwegian Research Council 
(NFR).

\end{multicols}

\newpage
\begin{table}
\caption{Summed TSC intensities $\sum_j{I^{\gamma\gamma}_{ijf}}$ over all
intermediate states $j$, and calculated under the assumption of $E1$ or $M1$
multipolarity of the soft dipole resonance.}
\label{tab:sum}
\begin{tabular}{ccccccccc}
nucleus&\multicolumn{2}{c}{$^{162}$Dy}&\multicolumn{2}{c}{$^{172}$Yb}
&\multicolumn{4}{c}{$^{161}$Dy}\\
$J_i^\pi\rightarrow J_f^\pi$&\multicolumn{2}{c}{$3^+\rightarrow 2^+$}
&\multicolumn{2}{c}{$0^-\rightarrow 2^+$}
&\multicolumn{2}{c}{$1/2^+\rightarrow 5/2^+$}
&\multicolumn{2}{c}{$1/2^+\rightarrow 5/2^-$}\\
multipolarity&$E1$&$M1$&$E1$&$M1$&$E1$&$M1$&$E1$&$M1$\\
$\sum_j{I^{\gamma\gamma}_{ijf}}$[\% per capture]
&4.9&4.2&8.1&11.3&4.2&2.7&0.9&2.0\\
\end{tabular}
\end{table}

\end{document}